\documentclass[11pt]{article}

\usepackage[final]{acl}
\usepackage{natbib}
\usepackage{booktabs}

\usepackage{times}
\usepackage{latexsym}

\usepackage[T1]{fontenc}

\usepackage[utf8]{inputenc}

\usepackage{microtype}

\usepackage{inconsolata}

\usepackage{graphicx}

\usepackage{enumitem}
\usepackage{subcaption} 

\newcommand{\figureEvolution}{
\begin{figure*}[!ht]
    \centering
    \includegraphics[width=\textwidth]{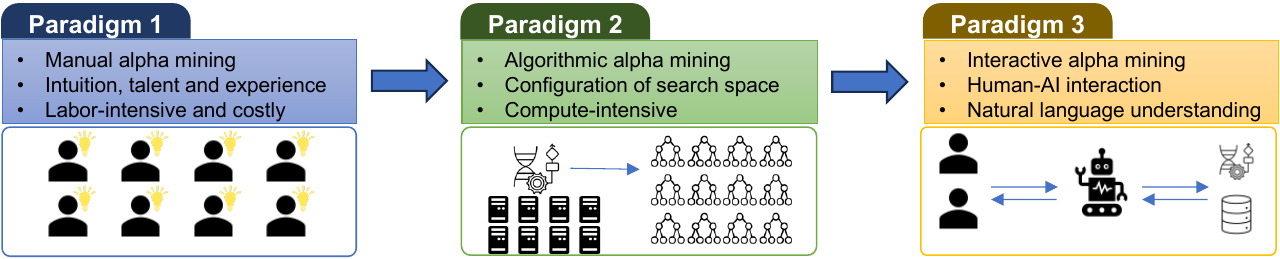}
    \caption{Evolution of alpha mining techniques.}
    \label{fig:evolution}
\end{figure*}
}

\newcommand{\figurePipeline}{
\begin{figure*}[!ht]
    \centering
    \includegraphics[width=\textwidth]{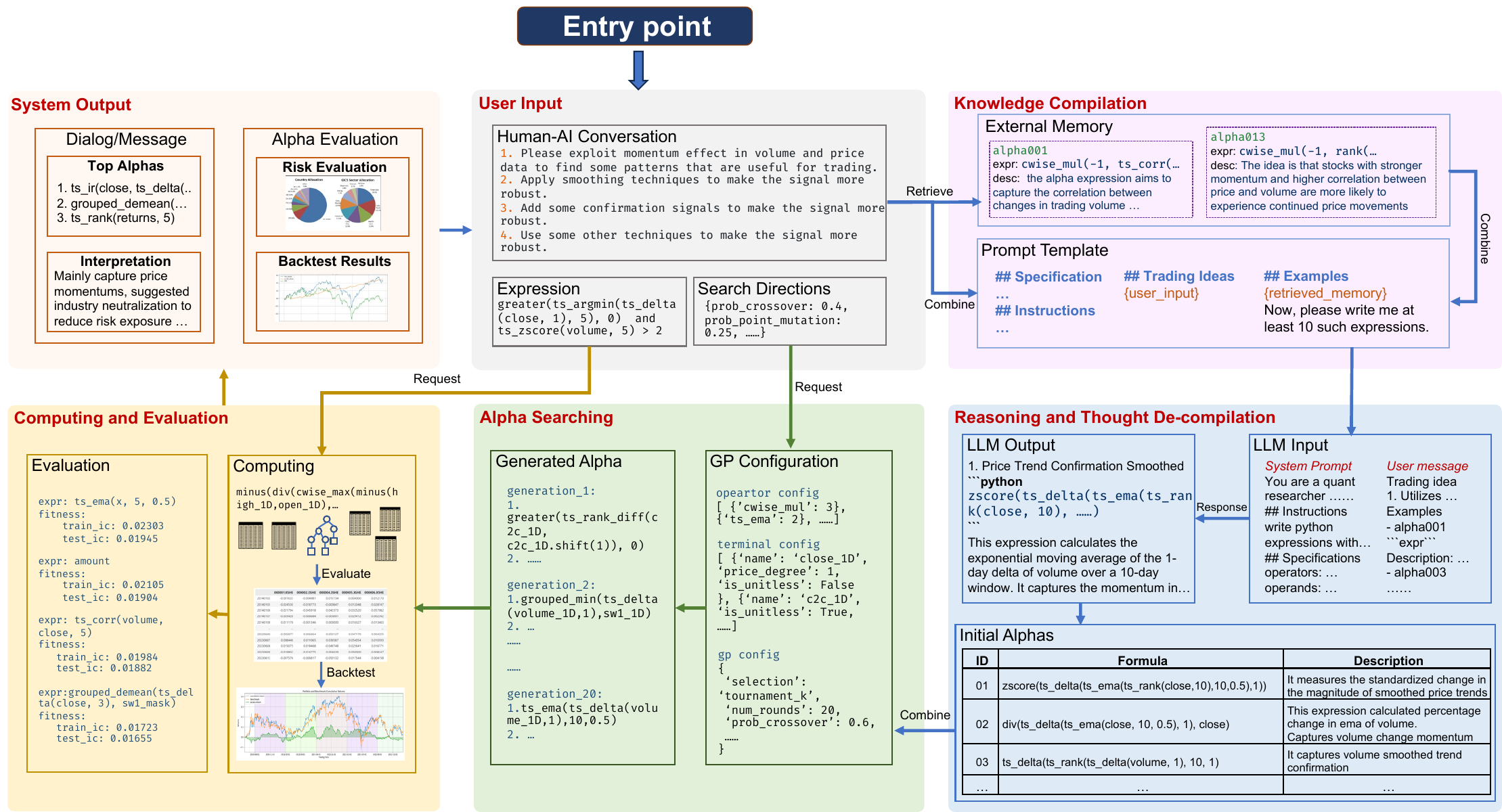}
    \caption{Alpha-GPT internal working pipeline: \normalfont{After a user inputs their ideas, the system goes into the knowledge compilation module. It uses external memory to pull similar examples, and combines them into the system prompt. The module passes everything to the LLM which creates valid alpha expressions and config files. These alphas are evaluated via Alpha Search, and results are presented to the user along with an interpretation provided by the Thoughts Decompiler.}}
    \label{fig:pipeline}
\end{figure*}
}

\newcommand{\figureKlineComp}{
\begin{figure*}
    \centering
    \begin{subfigure}{0.3\textwidth}
        \centering
        \includegraphics[width=\linewidth]{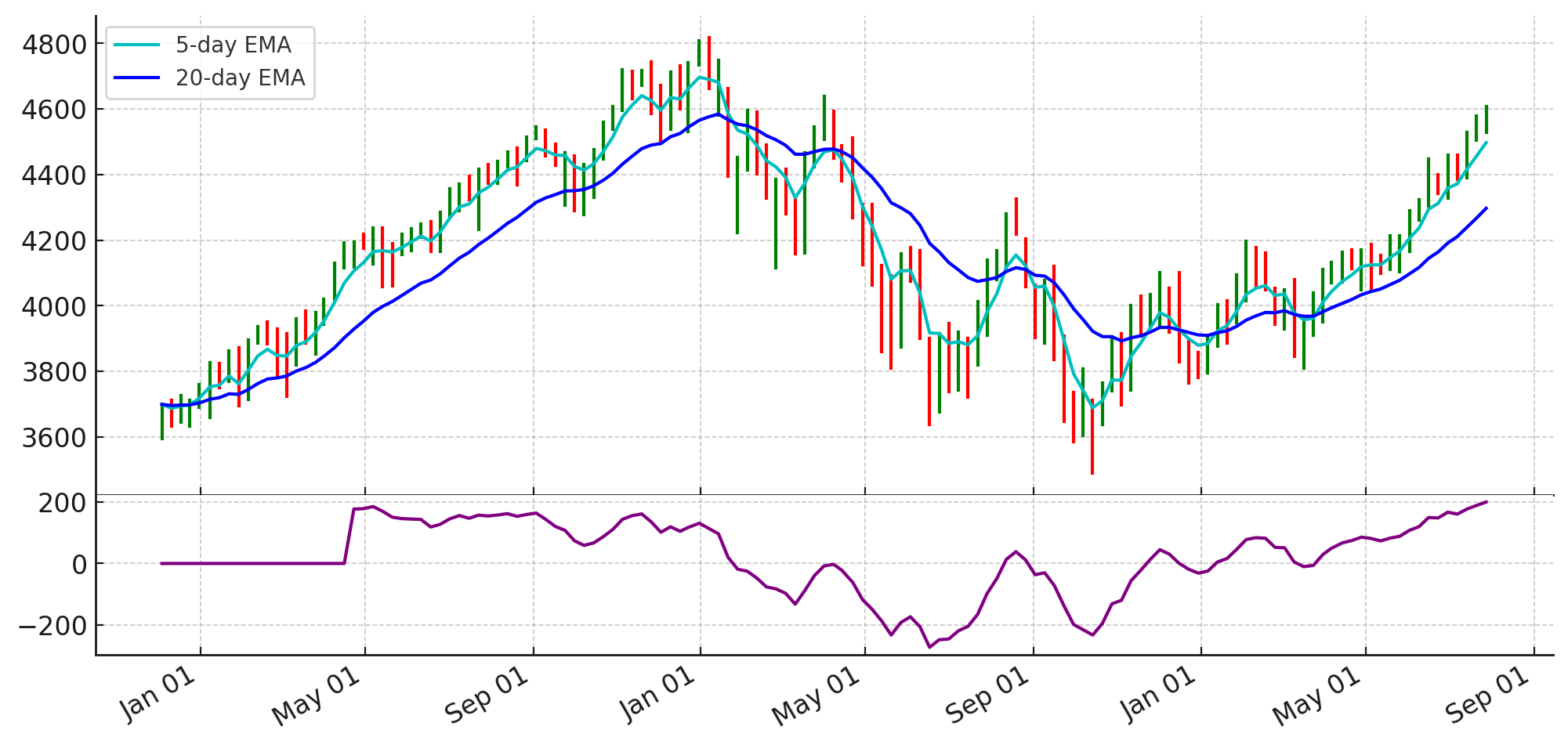} 
        \caption{Golden-cross pattern}
        \label{fig:subfigure1}
    \end{subfigure}
    \begin{subfigure}{0.3\textwidth}
        \centering
        \includegraphics[width=\linewidth]{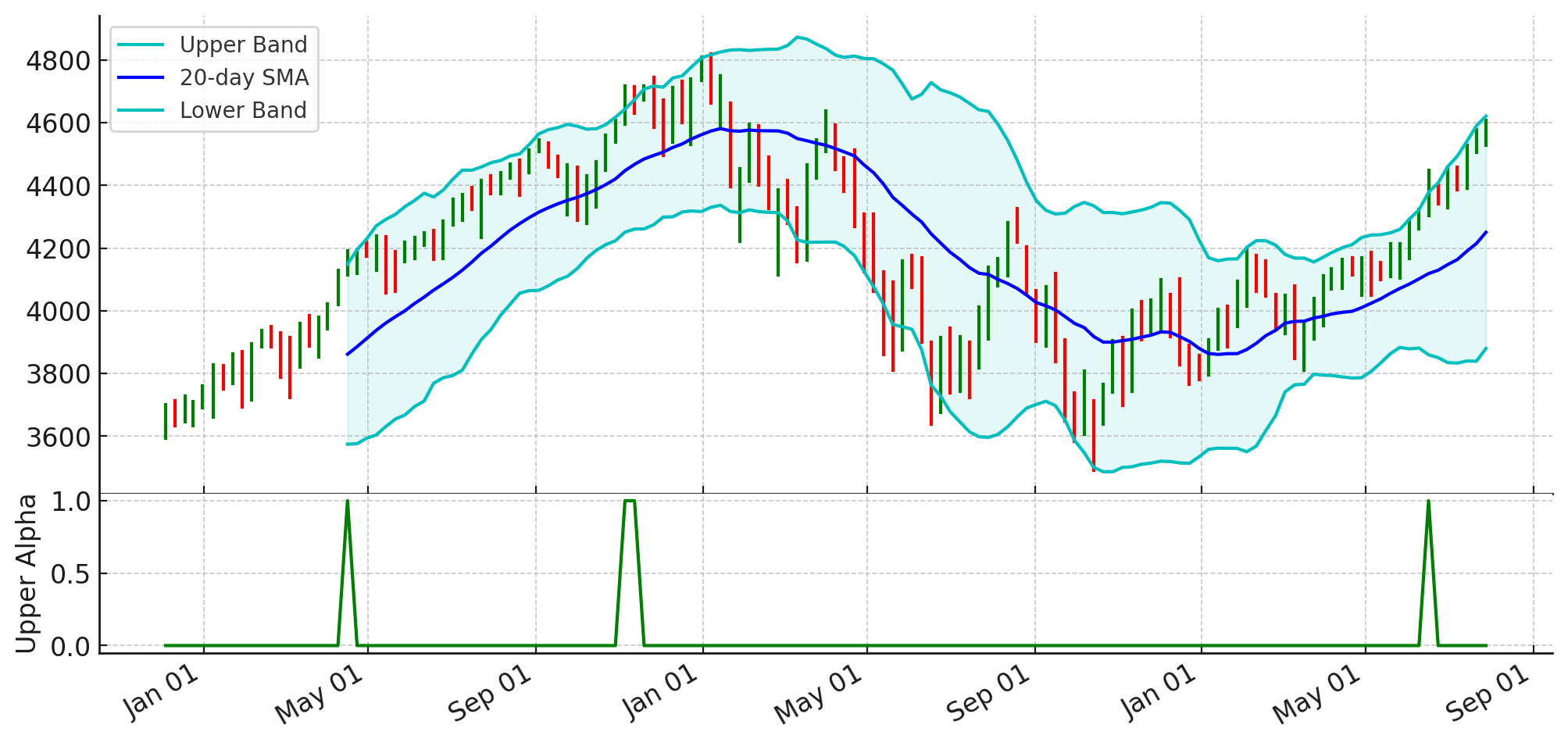} 
        \caption{Bollinger bands (upper breakout)}
        \label{fig:subfigure2}
    \end{subfigure}
    \begin{subfigure}{0.3\textwidth}
        \centering
        \includegraphics[width=\linewidth]{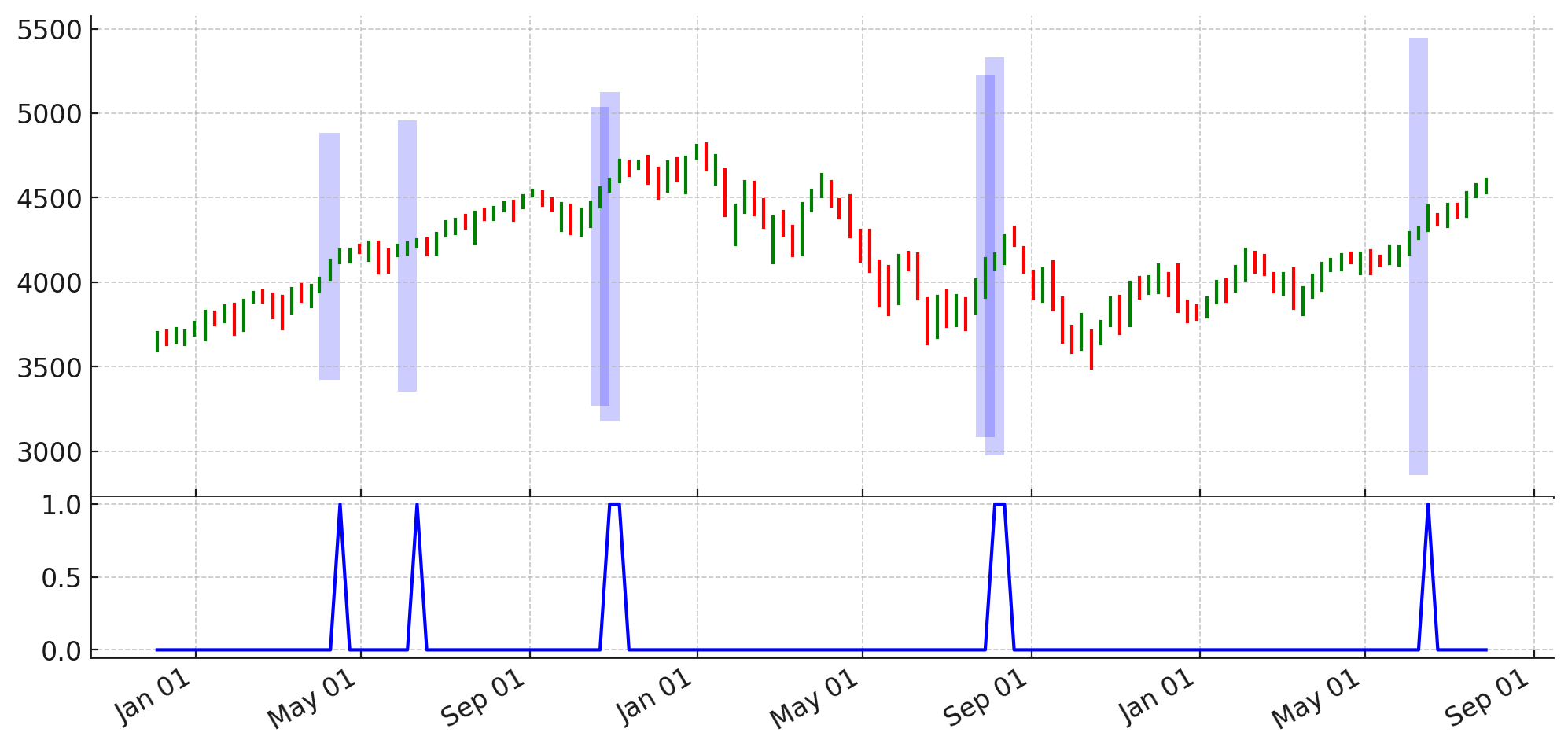} 
        \caption{Three white soldiers}
        \label{fig:subfigure3}
    \end{subfigure}
    \caption{Trading patterns and the corresponding alphas generated by Alpha-GPT that capture them.
    }
    \label{fig:kline_vis}
\end{figure*}
}

\newcommand{\figureAlphaExp}{
\begin{figure}[htbp]
    \centering
    \includegraphics[width=\linewidth]{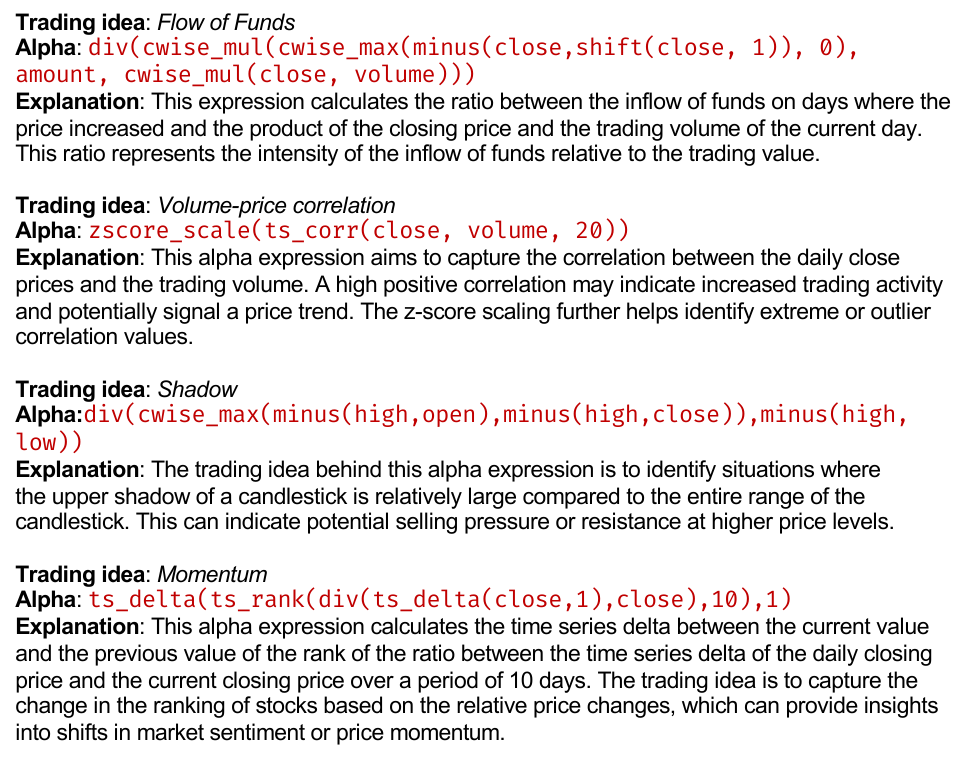}
    \caption{Alphas generated based on trading ideas and the corresponding explanations generated by Alpha-GPT.}
    \label{fig:alphaexp}
\end{figure}
}

\newcommand{\figureSearchEnhancement}{
\begin{figure}[!t]
        \centering
        \includegraphics[width=\linewidth]{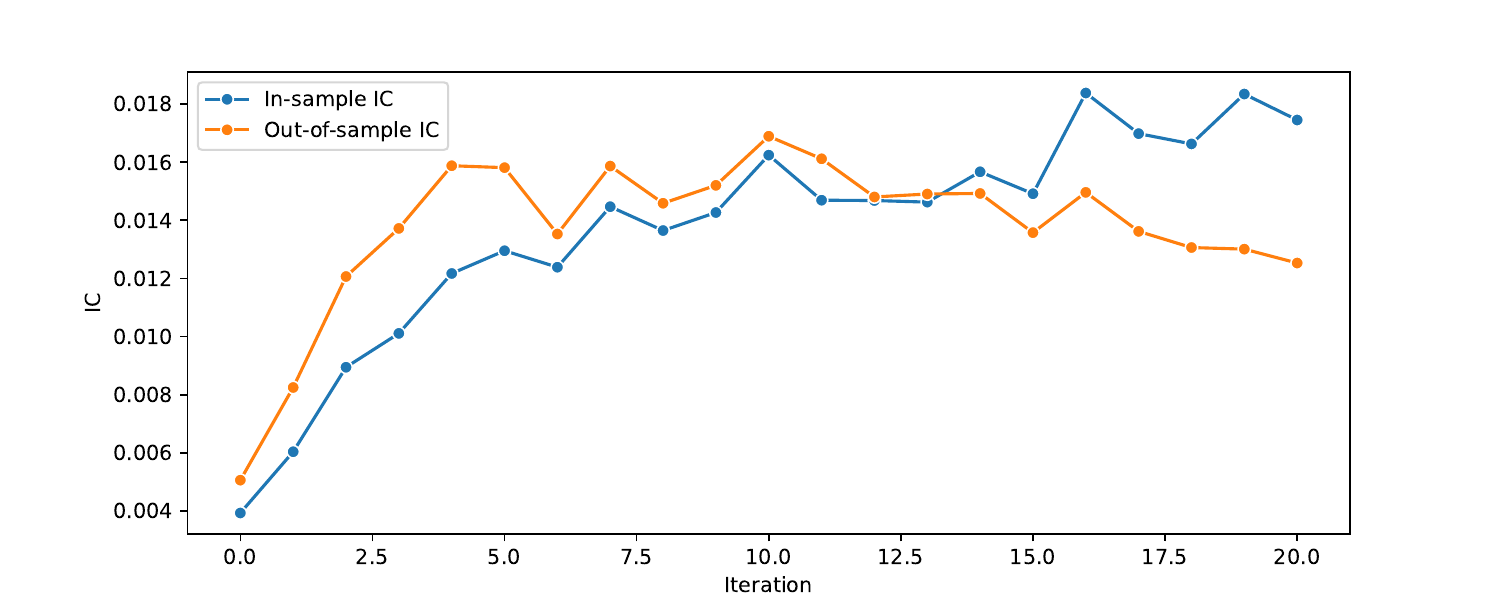}
        \caption{Search Enhancement curve}
        \label{fig:search_enh}
\end{figure}
}

\newcommand{\tableOperators}{
\begin{table}[!t]
  \centering
  \caption{Operators used in the experiment}
  \resizebox{\linewidth}{!}{
    \begin{tabular}{p{7em}|p{26.125em}}
    \toprule
    \multicolumn{1}{l|}{\textbf{Type}} & \multicolumn{1}{l}{\textbf{Operators}} \\
    \midrule
    time-series & shift,ts\_corr,ts\_cov, ts\_decayed\_linear, ts\_min, ts\_max, ts\_argmax, ts\_argmin, ts\_argmaxmin\_diff, ts\_max\_diff, ts\_min\_diff, ts\_mean, ts\_median, ts\_zscore\_scale, ts\_maxmin\_scale, ts\_skew, ts\_kurt, ts\_delta.ts\_delta\_ratio, ts\_ir, ts\_decayed\_linear, ts\_ema, ts\_percentile, ts\_linear\_reg, ts\_rank, ts\_sum, ts\_product, ts\_std, \\
    \midrule
    cross-sectional & zscore\_scale, winsorize\_scale, normed\_rank,cwise\_max, cwise\_min \\
    \midrule
    group-wise & grouped\_demean, grouped\_max, grouped\_min, grouped\_sum, grouped\_mean, grouped\_std, grouped\_zscore\_scale, grouped\_winsorize\_scale,  \\
    \midrule
    element-wise & relu, neg, abs, log, sign, pow,pow\_sign, round, add, minus, cwise\_mul, div,greater,less, normed\_rank\_diff \\
    \bottomrule
    \end{tabular}%
    }
  \label{tab:operator}%
\end{table}%
}

\newcommand{\tableRank}{
\begin{table}[!t]
  \centering
  \caption{Alpha-GPT's comparison with human-crafted factors}
    \begin{tabular}{c|ccc}
    \toprule
          & Return & Sharpe & MDD \\
    \midrule
    Top-1 (Human) & \textbf{21\%} & \textbf{6.88}  & \underline{1.61\%} \\
    Top-5\% & 16\% & 5.42  & \textbf{1.59\%} \\
    Top-10\% & 13\% & 4.16  & 3.58\% \\
    \midrule
    Alpha-GPT & \underline{14\%} & \underline{5.47}  & 2.36\% \\
    \bottomrule
    \end{tabular}%
  \label{tab:rank}%
\end{table}%
}

\newcommand{\tableConsistency}{
\begin{table}[!t]
  \centering
  \caption{Consistency comparison between a junior human researcher and Alpha-GPT}
    \begin{tabular}{c|cc}
    \toprule
    \multicolumn{1}{c|}{} & Score & Win rate \\
    \midrule
    Human & 6.81  & 13.40\% \\
    Alpha-GPT & 8.16  & 86.60\% \\
    \bottomrule
    \end{tabular}%
  \label{tab:consistency}%
\end{table}%
}

\newcommand{\tableIter}{
\begin{table}[!t]
  \centering
  \caption{Alpha IC comparison between different stages. ``Seed'' means the initial alpha generated by Alpha-GPT. ``SE'' means 10 rounds of search enhancement on the initial alpha. ``IT+SE'' means after 1 round of interaction and then 10 rounds of search enhancement.}
    \begin{tabular}{c|ccc}
    \toprule
    Alpha & Seed  & SE    & IT + SE \\
    \midrule
    IC    & 0.58\% & 1.23\% & 2.23\% \\
    \bottomrule
    \end{tabular}%
  \label{tab:iter}%
\end{table}%
}

\newcommand{\tableWorldQuantIQC}{
\begin{table*}[!t]
  \centering
    \caption{WorldQuant International Quant Championship 2024 Stage 2 Results (by June 25th, 2024)}
    \resizebox{\textwidth}{!}{
        \begin{tabular}{c|cccc}
        \toprule
              & Number of Qualified Alphas Generated & Total Score & In-sample Score & Out-of-sample Score \\
        \midrule
        Worldwide Top-1 & 103   & 52058 & 57899 & 50111 \\
        Worldwide Top-10 & 47    & 47112 & 42303 & 48715 \\
        Regional Top-1 & 91    & 50920 & 55890 & 49264 \\
        Regional Top-10 & 74    & 35999 & 26292 & 39325 \\
        Alpha-GPT & 81    & 48866 & 65505 & 43319 \\
        \bottomrule
        \end{tabular}%
    }
  \label{tab:iqc_results}%
\end{table*}%
}

\title{Alpha-GPT: Human-AI Interactive Alpha Mining for Quantitative Investment}

\begin{document}

\author{
    Saizhuo Wang$^{1*}$, Hang Yuan$^{1*}$, Leon Zhou$^3$, Lionel M. Ni$^{1,4}$, \\ \textbf{Heung-Yeung Shum}$^{1,2}$, \textbf{Jian Guo}$^2$ \\
    $^1$ HKUST, $^2$ IDEA Research, $^3$ Columbia University, $^4$ HKUST-GZ \\
    \texttt{\{swangeh, hyuanak\}@connect.ust.hk, leon.zhou@columbia.edu,}\\ 
    \texttt{ni@ust.hk, \{hshum,guojian\}@idea.edu.cn} \\
    \small $^*$Equal contribution
}

\maketitle
\begin{abstract}
One of the most important tasks in quantitative investment research is mining new alphas (effective trading signals or factors). Traditional alpha mining methods, either hand-crafted factor synthesis or algorithmic factor mining (e.g., search with genetic programming), have inherent limitations, especially in implementing the ideas of quant researchers. In this work, we propose a new alpha mining paradigm by introducing human-AI interaction, and a novel prompt engineering algorithmic framework to implement this paradigm by leveraging the power of large language models. Moreover, we develop Alpha-GPT, a new interactive alpha mining system framework that provides a heuristic way to ``understand'' the ideas of quant researchers and outputs creative, insightful, and effective alphas. We demonstrate the effectiveness and advantage of Alpha-GPT via a number of alpha mining experiments. In particular, we evaluated Alpha-GPT's performance in the \textbf{WorldQuant International Quant Championship 2024}, where it demonstrated results comparable to those of top-performing human participants, ranking among \textbf{top-10} over 41000 teams worldwide.
These findings suggest Alpha-GPT's significant potential in generating highly effective alphas that may surpass human capabilities in quantitative investment strategies.
\end{abstract}

\section{Introduction}
A trading alpha \cite{tulchinsky_introduction_2019} is a financial signal or a function with predictive power over excess return or risk, and they are usually expressed via symbolic rules or formulas (machine learning alphas are getting more popular recently but they are not discussed in this work). Alphas play a vital rule in trading economics, and most research work in quantitative investment focuses on how to find good alphas. See \cite{kakushadze_101_2016} for a number of such formulaic alphas (e.g., $-\frac{close-open}{(high-low) + 0.001}$ computes the increase from open price to close price relative to the intraday volatility, and the negative sign indicates a potential mean-reversion effect).

\figureEvolution
Traditionally, alpha mining has two paradigms (Figure \ref{fig:evolution}).
The first paradigm relies on manual modeling. Quant researchers attempt to translate their ideas/intuitions about financial markets into formulaic alphas, test their effectiveness and significance through backtest experiments, and analyze the reasons of success and failure. Usually, this process is repeated for many rounds to improve the performance of alphas. The success of this paradigm depends heavily on the talent and expertise of individuals and suffers from the problems of inefficiency and labor cost. On the other hand, the second paradigm seeks alphas through search algorithms such as genetic programming \cite{zhang_autoalpha_2020}. Since the search space, composed of all possible combinations for hundreds of operators and operands (features), is incredibly large, it is extremely compute-intensive to find satisfactory alphas during the alpha search process. 

Both of these paradigms exhibit common shortcomings. Firstly, it is a difficult process to find a precise and concise formulaic expression that encapsulates one's ideas about trading signals or observed trading opportunities and patterns. Examples include the formulaic representation of technical analysis patterns such as Ascending Triangles \cite{lo_foundations_2000} and Elliott Wave Theory \cite{elliott_rn_2005}, which exist but are hard to discover. Secondly, understanding and interpreting a large number of alphas selected by search algorithms is especially time-consuming and labor-intensive. Lastly, it is unreasonable to expect creative and effective alphas to come from the stroke-of-genius by researchers or the brute-force search by algorithms, but rather, it often comes from a repeated process of experimentation-and-analysis. However, designing and modifying the parameters and search configurations of algorithmic alpha mining systems is usually a menial task for researchers. 

To address these challenges, we propose the third alpha mining paradigm which enhances human-AI interaction to improve the effectiveness and efficiency of alpha research. Based on this new paradigm, we propose the architecture of an interactive alpha mining system, termed \textbf{Alpha-GPT}.
This system incorporates large language models (LLM) as a mediator between quantitative researchers and alpha search. Alpha-GPT has three key advantages. First, it can interpret users' trading ideas and translate them into fitting expressions, thanks to LLM's great natural language understanding and instruction-following capability. Secondly, Alpha-GPT can quickly understand, exploit and summarize top-performing alphas and meta-data via their natural/formal language expression, leveraging LLM's broad prior knowledge obtained via pretraining. Finally, the user can then suggest modifications to the alpha search, which the model will automatically make to future rounds of alpha mining, based on LLM's in-context learning and reasoning capabilities. This greatly simplifies the workflow (Figure \ref{fig:pipeline}) and allows the user to approach alpha mining from a high-level standpoint (in terms of abstract ideas).

Our contributions in this work can be summarized from these standpoints: (1) We define a new paradigm for alpha mining utilizing human-AI interaction to improve the effectiveness of alpha research. (2) We propose AlphaBot, an algorithm with domain knowledge compilation and decompilation methods to employ the LLM as a mediator for human-AI interaction. (3) We develop Alpha-GPT, a system to realize our proposed paradigm and a tool for quantitative researchers.  

\begin{figure*}
    \centering
    \includegraphics[width=\textwidth]{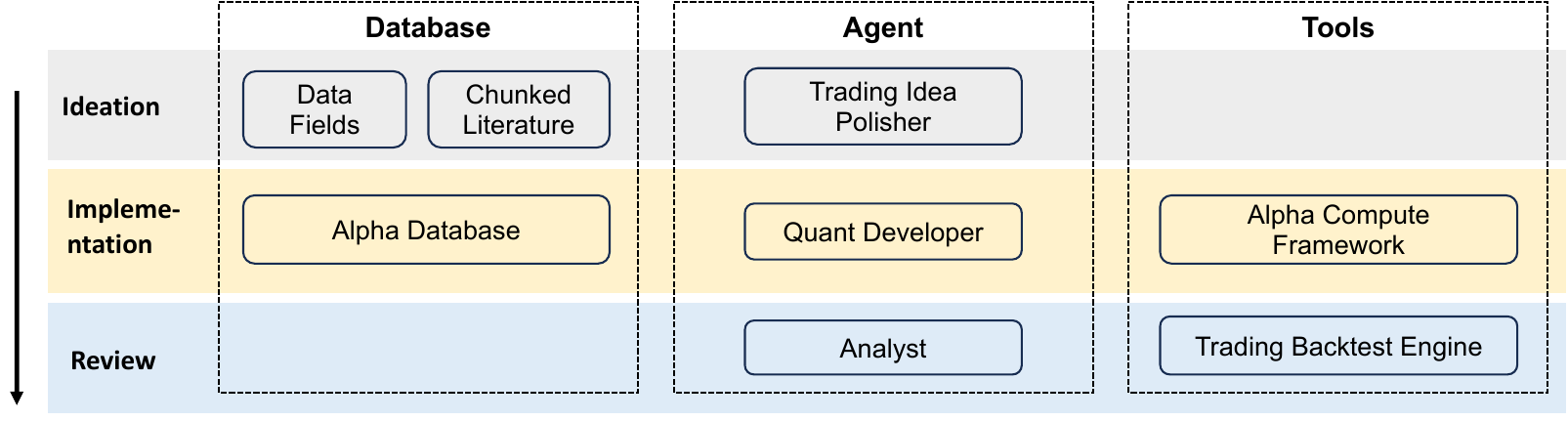}
    \caption{The agentic workflow of Alpha-GPT}
    \label{fig:agentic_workflow}
\end{figure*}

\section{Agentic Workflow}
\label{sec:workflow}
Taking inspiration from the established process of human quantitative researchers, Alpha-GPT employs an agentic workflow to generate and refine trading alphas. As illustrated in Figure \ref{fig:agentic_workflow}, this workflow is structured as an iterative process comprising three distinct stages: ideation, implementation, and review.

\subsection{Ideation}
The workflow is initiated in the \textbf{ideation} stage, wherein a quantitative researcher articulates a trading idea or market intuition using natural language. The principal agent in this stage is \textit{Trading Idea Polisher}, whose primary goal is to formalize the researcher's nascent idea into a structured prompt suitable for machine processing. To accomplish this, the agent queries a \textbf{Database} containing a corpus of literature and detailed specifications of available data fields. By leveraging this external knowledge base, the \textit{Trading Idea Polisher} augments the original query, disambiguates financial terminology, and incorporates contextual examples to ensure the precise capture of the user's intent.

\subsection{Implementation}
During the \textbf{implementation} stage, the refined idea from the preceding stage is operationalized into executable alpha expressions. The \textit{Quant Developer} agent, which leverages a Large Language Model (LLM), processes the structured prompt to generate a set of initial ``seed'' alpha expressions. These mathematical formulations are intended to embody the specified trading concept and are cataloged in the \textbf{Alpha Database}. Following this, the \textbf{Alpha Compute Framework} employs algorithmic search enhancement methods, notably genetic programming, to iteratively evolve and improve this initial set of alphas. This process yields a more diverse and sophisticated population of candidate alphas optimized for performance.

\subsection{Review}
In the terminal stage, \textbf{review}, the candidate alphas undergo rigorous empirical evaluation. The \textit{Analyst} agent coordinates this process, utilizing the \textbf{Trading Backtest Engine} as its primary analytical tool. This engine executes historical simulations to assess alpha performance against market data, generating quantitative metrics that include backtest returns, Information Coefficient (IC), and Sharpe ratio. The \textit{Analyst} agent then synthesizes these outputs, providing natural language summaries and interpretations of the top-performing alphas to the researcher. This interactive feedback loop enables the researcher to provide further direction for subsequent rounds of alpha mining, fostering a collaborative human-AI discovery process.

\begin{figure*}
    \centering
    \includegraphics[width=\textwidth]{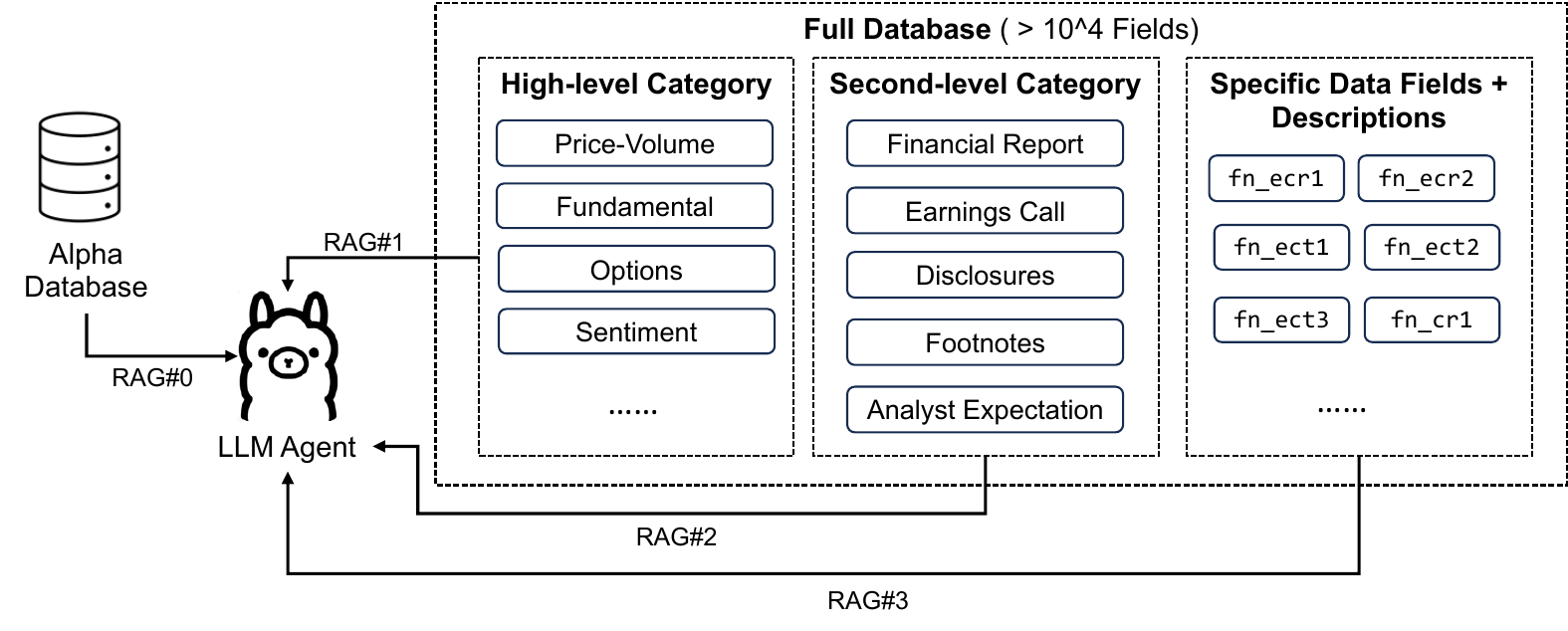}
    \caption{Alpha-GPT's hierarhical RAG in autonomous mode for large-scale quant database.}
    \label{fig:hierarchical_rag}
\end{figure*}

\section{Modes of Operation}
As a practical assistant tool for quantitative research, Alpha-GPT is designed to operate in two distinct modes: \textit{interactive mode} and \textit{autonomous mode}. 
In interactive mode, the system functions as a collaborative partner, where human researchers provide input and guidance throughout the agentic workflow. This approach is predicated on the recognition that human domain expertise and intuition in trading and investment often surpass the current capabilities of LLMs. In contrast, the autonomous mode enables the system to generate and iterate upon trading ideas independently. This mode is particularly useful when faced with exceptionally large quantitative databases, where it can perform a rapid and reliable bootstrap of satisfactory alphas that human researchers can subsequently analyze and develop further.

\subsection{Interactive Mode}
In the interactive mode (an example pipeline shown in Figure \ref{fig:pipeline}), Alpha-GPT serves as an intelligent interface that bridges the gap between a researcher's conceptual ideas and their empirical validation. The human researcher remains central to the discovery process, initiating the workflow by providing trading ideas in natural language and offering feedback at the review stage of each iteration. In this collaborative paradigm, Alpha-GPT acts as a co-pilot responsible for translating these abstract concepts into precise, formulaic alpha expressions. It then manages the computationally intensive tasks of alpha enhancement using methods like genetic programming and executes rigorous backtesting for performance evaluation. Finally, the system synthesizes the complex results into comprehensible natural language summaries, facilitating human review and decision-making to guide the next cycle. This synergy between human intuition and the system's advanced computational capabilities serves to accelerate the research cycle.

\subsection{Autonomous Mode}
The autonomous mode is engineered for the systematic exploration of large-scale quantitative databases, which can contain tens of thousands of data fields. In such scenarios, providing the complete documentation of all available data to an LLM would overwhelm its context window, both in terms of token limits and information density. To surmount this challenge, Alpha-GPT employs a hierarchical Retrieval-Augmented Generation (RAG) strategy, as depicted in Figure \ref{fig:hierarchical_rag}. This strategy enables the LLM agent to autonomously discover novel trading ideas by navigating the database in a structured, top-down manner.

The process commences with the LLM agent analyzing the existing \textbf{Alpha Database} to learn the characteristics of previously successful alphas (RAG\#0). Guided by this initial analysis, the agent then queries the \textbf{High-level Categories} of the full database, such as `Price-Volume` or `Sentiment`, to identify broad, promising domains for new alpha discovery without retrieving excessive detail (RAG\#1). Following this, the agent performs a more focused query on the corresponding \textbf{Second-level Categories}, like `Earnings Call`, to progressively narrow the search space (RAG\#2). In the final step, the agent retrieves the detailed descriptions for \textbf{Specific Data Fields} within the chosen sub-category, and armed with this granular information, it can formulate a novel, concrete trading idea and generate the associated alpha expression (RAG\#3). This hierarchical framework allows Alpha-GPT to methodically explore a vast and complex feature space, effectively managing context size while continuously generating novel ideas.

\section{System Architecture}
The overall system architecture of Alpha-GPT is illustrated in Figure \ref{fig:architecture}. It is a multi-layered framework composed of a user-facing interface, a core LLM agent, an algorithmic mining engine, and a computation acceleration layer.

\subsection{WebUI and LLM Agent}
The top layers of the architecture facilitate human-AI interaction. The Web-based User Interface (WebUI) is the primary entry point for a quantitative researcher. It includes a \textbf{Dialog Box} for natural language interaction, a \textbf{Mining Session Manager} to organize distinct research threads, and an \textbf{Alpha Mining Dashboard} for comprehensive visualization of experiments and performance analytics.
The LLM Agent, termed as the \textbf{AlphaBot} layer, serves as the core intelligence of the system. It employs a standard prompt engineering pipeline to translate user intent into structured tasks. This process leverages Retrieval-Augmented Generation (RAG) over a vector database of financial literature and historical alphas to ground the model's outputs. The agent's responses are then processed through a structured output parsing and validation module to ensure the generation of syntactically correct and semantically valid alpha expressions for the backend systems.

\subsection{Backend Systems}

\paragraph{Algorithmic Alpha Mining}
This layer serves the search enhancement function in Alpha-GPT. It implements an algorithmic workflow by taking the seed alphas generated by AlphaBot and iteratively improving them based on received search commands and configurations. The layer consists of four modules. The \textbf{Alpha Search Enhancement} module uses techniques like genetic programming to generate a diverse set of alpha candidates. Qualified alphas are then filtered by the \textbf{Evaluation and Backtesting} module, which assesses performance against historical data. These alphas are further pruned and scored by the \textbf{Alpha Selection} module to remove redundancies and identify the most valuable signals. Finally, the \textbf{Alpha Deployment} module prepares the finished alphas for live trading, ensuring the smoothness and correctness of real-time computation.

\paragraph{Alpha Computation Acceleration}
Alpha computation requires processing vast amounts of financial data, and the computational overhead of handling high-frequency data makes acceleration a key requirement. The alpha computation acceleration layer employs several key techniques to meet these demands, including the use of streaming algorithms for rolling window computations, vectorized computation to leverage hardware concurrency, SIMD/SIMT instructions for parallel data processing, memory optimization techniques like pre-allocation, and GPU acceleration for data-intensive tasks.

\section{Evaluations}

In order to assess the impact of Alpha-GPT on enhancing researchers' productivity in identifying relevant factors, we carry out a combination of quantitative and qualitative studies. The quantitative experiments aim to validate the effectiveness of Alpha-GPT by evaluating its performance based on given sets of trading ideas or databases, while the qualitative experiments (Section \ref{sec:additional_qualitative}) aim to showcase successful instances of its application.
The results below are intended to verify the following questions:  
\textbf{(1)} Can Alpha-GPT improve quant research efficiency via human-AI interaction?  
\textbf{(2)} Can the algorithmic search enhancement module improve the quality of generated alpha?  
\textbf{(3)} Can Alpha-GPT ultimately lead to better alphas?

\subsection{Experimental Setup}
\label{sec:exp_setup}
Without further specifications, the experiments below are conducted with the following setups.
\paragraph{Data and operators} We use intraday volume-price data of Chinese and US stocks. The data include the basic candlestick chart data (OHLCV), volume-weighted average price (VWAP), and sector data. The operators we use include 19 basic operators implemented in \cite{guo_hxpy_2023} including time-series operations, cross-sectional operations, group-wise operations and basic element-wise operations, as shown in Table \ref{tab:operator}. Besides, we also incorporated operators from existing libraries such as scipy and torch.
\tableOperators

\paragraph{Knowledge Library} 
We construct the knowledge library based on the alphas proposed in \cite{kakushadze_101_2016} and a proprietary alpha base. For each alpha, we first decompose it into sub-expressions and explain them. Then we explain the combination of these sub-expressions to form the whole trading idea. Document embeddings are indexed via Faiss\footnote{https://github.com/facebookresearch/faiss}. Note that we only employed external memory when generating alphas for trading ideas that align well with those in the alpha base. Importantly, the knowledge library serves as an auxiliary resource to enhance interpretability and consistency, rather than as a source of direct alpha reuse. Alpha-GPT remains capable of producing novel alphas beyond the scope of the library, and our experiments confirm that a large portion of generated alphas are not present in either the literature or the proprietary base. The inclusion of these resources thus does not compromise novelty but instead provides grounding and domain context for the generation process.

\paragraph{LLM and Adapter} We used Llama3 70B  \cite{grattafiori_llama_2024} as the chat model and BGE-M3 \cite{chen_bge_2024} as the embedding model.

\tableWorldQuantIQC

\subsection{Efficiency Improvement}
We evaluate Alpha-GPT's ability to improve research efficiency by assessing its effectiveness in translating trading ideas into alphas and its capacity to develop stronger alphas through iterative refinements.

\paragraph{Translation Consistency}
\tableConsistency
To verify Alpha-GPT's ability to enhance researchers' efficiency by providing accurate and high-quality factors, we conducted a comparative study. We collected generated alphas based on a trading idea dataset from both Alpha-GPT and a group of human quant researchers. The human group comprised five quant researchers with experience ranging from 0.5 to 2 years. The trading idea dataset was randomly split into five parts, with each human researcher tasked with writing alphas based on a specific split.
For evaluation, we prompted GPT-4 to score the generated alphas on a scale of 1 to 10 (with 10 being the highest score) and select the superior one. The average results are presented in Table \ref{tab:consistency}. The results show that the factors generated by Alpha-GPT consistently outperformed those produced by human researchers. This outcome strongly indicates Alpha-GPT's effectiveness in improving research efficiency by accurately translating trading ideas into high-quality factors.
This experiment demonstrates Alpha-GPT's potential to significantly enhance the productivity of quant research teams, particularly in the crucial task of transforming conceptual trading ideas into concrete, implementable factors.

\paragraph{Human-AI Iterative Refinement}
We also verify the effectiveness of Alpha-GPT in helping improve alpha research in through human-AI interaction. We first simulated a human user using another LLM (GPT-4) with specifically designed prompts. For each trading idea in the dataset, this simulated human user will send it to Alpha-GPT and interact with it for another round, based on the explanation generated by Alpha-GPT in the first round. Then, we evaluate the IC of the factors that are generated initially, after search enhancement, and after 1 round of interaction \& search enhancement. The result is shown in Table \ref{tab:iter}, where consistent improvements in factor IC demonstrates the effectiveness of interaction.
\tableIter

\subsection{Search Enhancement}
\figureSearchEnhancement
To validate the effectiveness of the alpha mining layer in consistently enhancing factors, we analyzed the information coefficient (IC) of alphas generated through multiple rounds of search enhancement, both in-sample and out-of-sample. Figure \ref{fig:search_enh} illustrates the IC curves over 20 iterations, revealing several key insights. Both in-sample and out-of-sample ICs show a sharp initial increase (iterations 0 to 5), indicating rapid improvement of initial factors. The in-sample IC (blue line) demonstrates a general upward trend throughout, suggesting continuous factor enhancement on training data. Notably, the out-of-sample IC (orange line) stabilizes after the initial rise, indicating that improvements generalize well to unseen data and mitigating overfitting concerns. Both curves appear to converge around the 15th iteration, suggesting an optimal stopping point for the enhancement process.

\subsection{Stronger Alphas}
To evaluate Alpha-GPT's ability to generate superior alphas and investment strategies, we designed an automated testing scenario simulating collaboration between human researchers and AI. We created a meta-database of operands (data fields) and operators with detailed descriptions. A specially prompted LLM was then used to systematically explore these fields and generate alphas with strong performances, simulating a human researcher interacting with Alpha-GPT to search for high-performing alphas. This process incorporates elements similar to traditional methods such as genetic programming, but with the search guided by the LLM.

\paragraph{High-frequency Trading Competition}
\tableRank
We evaluated Alpha-GPT in following the same evaluation protocol of a concluded alpha competition in high-frequency trading.\footnote{Joinquant alpha competition \href{https://www.joinquant.com/view/competition/highfrequency}{\textbf{link}}} Specifically, we incorporated self-improving mechanism \cite{wang_quantagent_2024} to generate factors and compared the result with human leaderboards, as shown in Table \ref{tab:rank}. It can be seen that Alpha-GPT achieved Top 5\%-10\% performance of human participants. Notably, the initial competition duration was one month, but Alpha-GPT was able to reach a comparable performance level in just one day.

\paragraph{WorldQuant International Quant Competition}
For a more practical and challenging scenario, we deployed our automation to the WorldQuant International Quant Championship (IQC) 2024 \footnote{WorldQuant IQC \href{https://www.worldquant.com/iqc/}{\textbf{link here}}.}, the premier competition in formulaic alpha mining that involves more than 41,000 participants from over 100 countries and 5,000 universities. The competition offers a vast exploration space with over 5,000 operand data fields spanning price-volume, fundamentals, derivatives, news sentiment, and more, along with over 100 operators of various types. The platform applies strict criteria for alpha qualification, considering factors such as alpha return, turnover, and Sharpe ratio. Importantly, our evaluation was conducted in real time during the official competition period, ensuring that no future information leakage occurred. 
As presented in Table \ref{tab:iqc_results}, the results demonstrate that our automated Alpha-GPT system can generate performant alphas, ranking among the top 10 worldwide and top 3 regionally. In particular, Alpha-GPT produces a comparable number of qualified alphas to top human competitors and achieves the highest in-sample score. The system's out-of-sample score is also highly competitive, indicating that alphas generated based on the LLM's prior knowledge generalize well and possess strong underlying logic. These impressive results underscore Alpha-GPT's potential to achieve superhuman performance in alpha mining.

\subsection{Qualitative Results}
\label{sec:additional_qualitative}
\paragraph{Idea-Formula Consistency}
We demonstrate that Alpha-GPT can generate formulaic alphas that are consistent with the user's given trading idea. 
Figure \ref{fig:kline_vis} illustrates the generated alpha expressions based on given trading ideas and their correspondence to the patterns in the candlestick chart. The candlestick chart is plotted from the weekly data of the S\&P500 index from 2020 to 2023.
The first trading idea aims to capture the divergence of two moving average prices with differing lookback windows and the generated factor successfully reflects this curve. The second trading idea characterizes the breakout signals of Bollinger bands, and the corresponding alpha is a binary signal that gets activated when the upper bound is crossed. The third trading idea aims to capture three consecutive bullish movements on the candlestick chart, and the generated alpha successfully identified those patterns.
These examples demonstrate that the generated alphas correctly capture the trading ideas.

\paragraph{Alpha Explanation}
Figure \ref{fig:alphaexp} presents examples of alpha expressions generated by Alpha-GPT based on given trading ideas, and the corresponding natural language explanations of these alphas also generated by Alpha-GPT. From these examples we can see that Alpha-GPT can provide appropriate explanations of the generated alphas, relieving the burden of human researchers to interpret these expressions by themselves.

\section{Related Work}
A lot of algorithms have been studied for formulaic alpha mining. Examples include Monte Carlo random search, Markov-chain Monte Carlo \cite{jin_bayesian_2020}, genetic programming \cite{cui_alphaevolve_2021} and their variants \cite{zhang_autoalpha_2020}, and reinforcement learning \cite{yu_generating_2023}.
However, these methods all require the user to directly define the algorithmic configurations, providing limited interactivity compared with Alpha-GPT.
Meanwhile, LLMs such as GPT \cite{brown_language_2020} have demonstrated emergent abilities \cite{wei_emergent_2022} and achieved superior performance on various tasks.
Besides, LLMs have also shown great reasoning \cite{wei_chain--thought_2022,yao_tree_2023} and planning capabilities \cite{yao_react_2022}.
In this way, an LLM can be regarded as a core thinking module and be integrated with various peripheral tools \cite{schick_toolformer_2023} to form intelligent LLM-powered agents \cite{weng_llm_2023}.


\newpage
\bibliography{references}

\newpage
\appendix

\figurePipeline

\begin{figure*}[!ht]
    \centering
    \includegraphics[width=\textwidth]{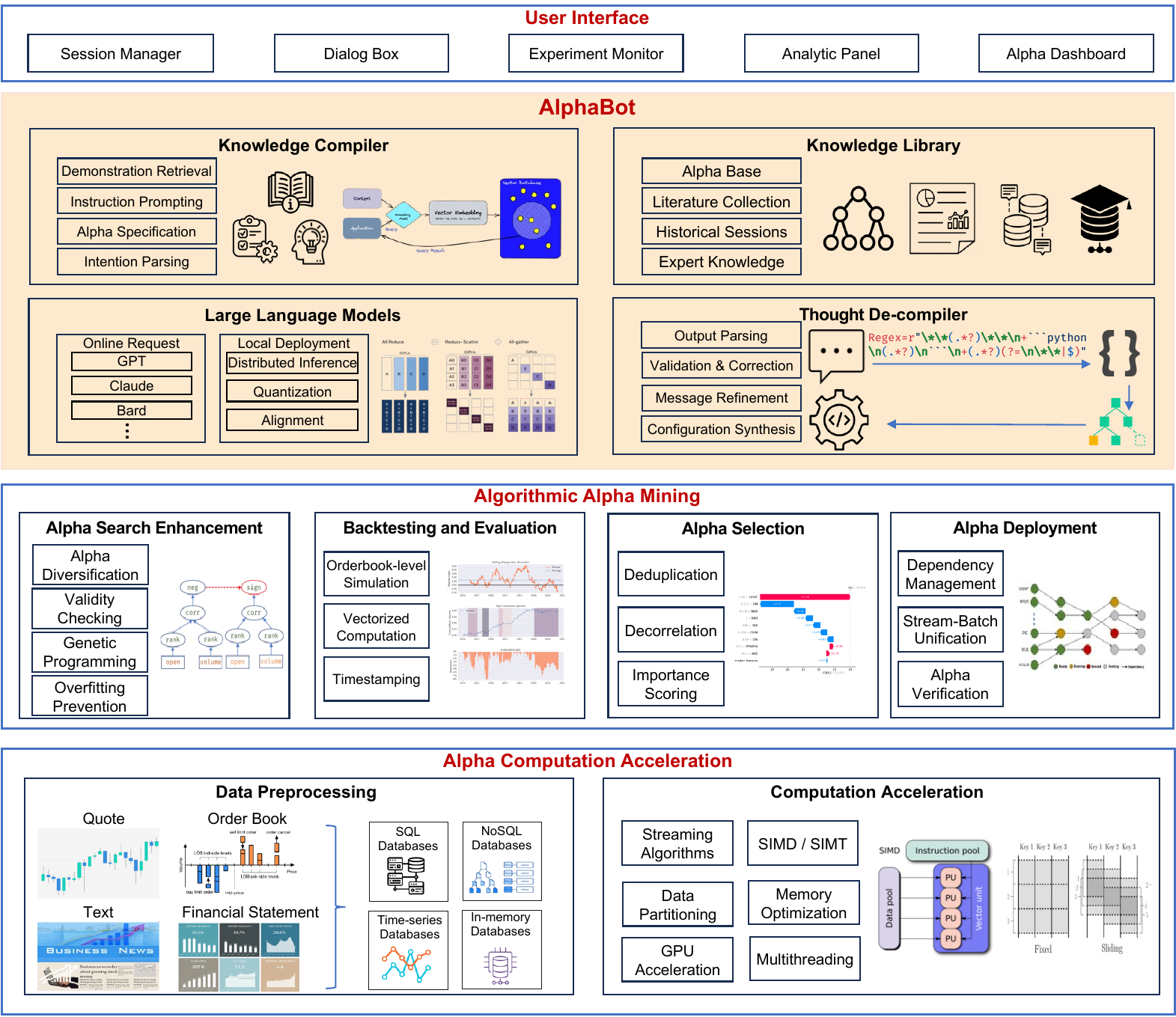}
    \caption{Alpha-GPT system architecture. The AlphaBot layer is the key contribution of this paper and the lower-level modules is integrated from our existing systems. Part of this figure is cited from \protect\cite{guo_quant_2022,wu_how_2021,noauthor_fully_2021,noauthor_slundbergshap_nodate}.}
    \label{fig:architecture}
\end{figure*}

\figureKlineComp

\figureAlphaExp

\end{document}